# Multiple Muon Measurements with MACRO[*]

The MACRO Collaboration[†]

Presented by J. T. Hong

Department of Physics, Boston University, Boston, Massachusetts 02215, U.S.A.

**Abstract**

The MACRO experiment at Gran Sasso provides means for detailed studies of multiple coincident penetrating cosmic ray muons. In this paper we concentrate on the studies of the ultrahigh energy primary cosmic ray composition using muon bundle multiplicities, muon pair lateral and angular separation distributions.

## 1. INTRODUCTION

The chemical composition and energy spectra of primary cosmic rays bombarding the top of the Earth's atmosphere contain information about the origin of cosmic rays, providing constraints on the modeling of acceleration and propagation in the interstellar medium. Experiments with balloons and satellites provide direct measurements of the composition and energy spectra of the primary species up to energies of several hundred TeV. At higher energies, direct measurements require impractically large detectors and/or long exposure time due to the steeply falling energy spectra. Consequently, very little is known about the composition of primary cosmic rays at energies greater than several hundred TeV [1]. However, indirect information on their composition can be extracted from experiments with secondary particles from atmospheric cascades, such as surface measurements of extensive air showers or underground studies of penetrating high energy muons.

The rates of underground muon bundles of different multiplicities, as well as their lateral and angular separations, are sensitive to the chemical composition and energy spectra of the primary cosmic rays, above a threshold determined by the rock overburden ($\sim 50$ TeV at Gran Sasso). This sensitivity arises from the fact that heavy nuclei tend to generate a larger yield of charged pions and kaons, in the forward fragmentation (high rapidity) region, than that of light nuclei. It is these pions and kaons which can decay into the penetrating muons observed deep underground.

In addition to the composition of cosmic rays, the properties of hadronic interactions at very high energies and in the high rapidity region will also affect the multiplicity and separation of muons observed underground. Interactions with higher overall cross sections will tend to cause primaries to interact higher in the atmosphere which can result in wider spread between secondary muons. Similarly, heavy nuclei will tend to interact higher in the atmosphere and will also result in larger muon spread. However, if an interaction tends to produce high multiplicity with little increase in transverse momentum, then

---

multiplicity will rise if the energy is sufficiently high (it can fall below a particular energy threshold) but there will be no increase in transverse momentum. Hence, there are a number of ways that the primary energy, composition, and interactions can influence the muon distributions. By studying these distributions and their correlations, various composition and interaction models can be constrained.

In this paper we report results from analyses [2] of the multiple coincident penetrating cosmic ray muon data collected by the MACRO detector. We have measured the muon bundle rate as a function of multiplicity as well as the muon pair lateral and angular separation distributions. These measured quantities are compared with Monte Carlo predictions for different primary cosmic ray composition models and interaction models.

## 2. THE MACRO DETECTOR

The MACRO (Monopole, Astrophysics, and Cosmic Ray Observatory) detector [3], a large area, deep underground detector, was fully completed recently at the Gran Sasso National Laboratories in central Italy. It consists of large liquid scintillator counters, limited streamer tubes, and plastic track-etch detectors, which offer three redundant techniques for its primary physics goals of monopole searches [4]. Its large size ($76 \times 12 \times 9$ m$^3$) and excellent tracking capability allow detailed studies of multiple coincident penetrating cosmic ray muons. The rock overburden has a minimum thickness of 3200 meters of water equivalent, setting the surface muon energy threshold of $\sim 1.4$ TeV. Above this energy, the probability for a muon to penetrate this minimum thickness is 30% or greater.

The full MACRO detector including its upper deck started to take muon data in March 1994. This paper uses data collected from the lower deck, which commenced data taking in June 1991. Consisting of six supermodules, the lower deck measures $76 \times 12 \times 4.8$ m$^3$. It is surrounded on all sides by planes of large liquid scintillator counters. The tracking system consists of ten horizontal layers of streamer tubes separated by 32 cm of crushed rock absorber. Each tube is 12 m long, $3 \times 3$ cm$^2$ in cross section, and utilizes a 100 $\mu$m anode wire and a graphite cathode. A two-dimensional readout is performed using signals from anode wires and external 26.5° stereo angle strips. Six additional vertical layers of streamer tubes cover each vertical side. The intrinsic angular resolution is 0.2° for muons crossing ten horizontal planes. Taking into account multiple coulomb scattering in the rock overburden, the overall angular resolution is estimated to be $\sim 1°$.

MACRO accumulates underground muon data at the rate of $\sim 6.6 \times 10^6$ events/live year. Approximately $\sim 4.0 \times 10^5$ events/live year exhibit multiple muon tracks with lateral separations up to 70 m and $\sim 1.6 \times 10^3$ events/live year have multiplicities of ten or greater. Combined with our ever improving ability to model cosmic ray showers, these data provides a unique opportunity to study ultrahigh energy cosmic rays.

## 3. COSMIC RAY COMPOSITION

We have studied the primary cosmic ray composition using the multi-muon event rate as a function of multiplicity. This analysis uses a data sample collected in 3295 hr live time with all six lower supermodules. This data sample contains $\sim 2.5 \times 10^6$ muon events, of which $\sim 1.5 \times 10^5$ are multiple muons. The event selection uses the criteria established in a previous analysis [5], which used data from only two supermodules. In the previous analysis, a considerable amount of visual scanning was performed to establish the actual multiplicity. In this analysis, we correct the reconstructed multiplicity using a GEANT-based [6] detector simulation program. The following physics and detector effects are taken into account: electromagnetic showering down to 500 keV, charge induction of the streamer signal onto the stereo strip, electronic noise, inefficiencies, and failures of the tracking algorithm especially for high multiplicity events, track shadowing at small separations, etc. The simulated data are used to calculate the correction factors for transforming the reconstructed multiplicities in the two projective readout views into an actual multiplicity. This allows an objective assignment of the muon multiplicity, reducing the systematic uncertainties in the previous analysis dominated by the scanning uncertainties. Figure 1 shows the muon bundle rates for the one, two, and six supermodule data samples. The increase of acceptance with detector size is reflected in the increase in muon rates and sampling of high multiplicity events.

The experimental data are compared with the results of full simulations of the primary interaction, the atmospheric cascades, the muon propagation in the rock, and the aforementioned detector response. Two different shower simulation codes are used: HEMAS [7] with the addition of nuclear fragmentation based on the semi-superposition model [8] and SIBYLL [9]. The simulations are described in more detail below in Section 4. This paper reports results based on the HEMAS simulations.

We considered three different primary composition models: light and heavy compositions used in our previous analysis [5], and a constant mass composition (CMC) with fixed spectral indices [10]. The light and heavy compositions are extreme models: at increasingly higher energies the light composition contains a large proton component while the heavy composition contains a large Fe component. However, the models are constrained to reproduce the known abundances and spectra directly measured at $\leq 100$ TeV and to agree with extensive air shower measurements at higher energies. Therefore a comparison of the muon experimental rates with the predictions of these models gives an indication of the sensitivity of the experiment to primary composition.

Figure 2 shows the calculated energy ranges of the primary cosmic rays at a detection efficiency of 90% as functions of detected multiplicities for light and heavy composition models. The plots demonstrate that the primary energies explored this way increase with muon multiplicity. In particular, events with detected multiplicity $\gtrsim 10$ originate from primaries in an energy region entirely above the "knee", the steepening point of the cosmic ray energy spectra at $\sim 100$ TeV.

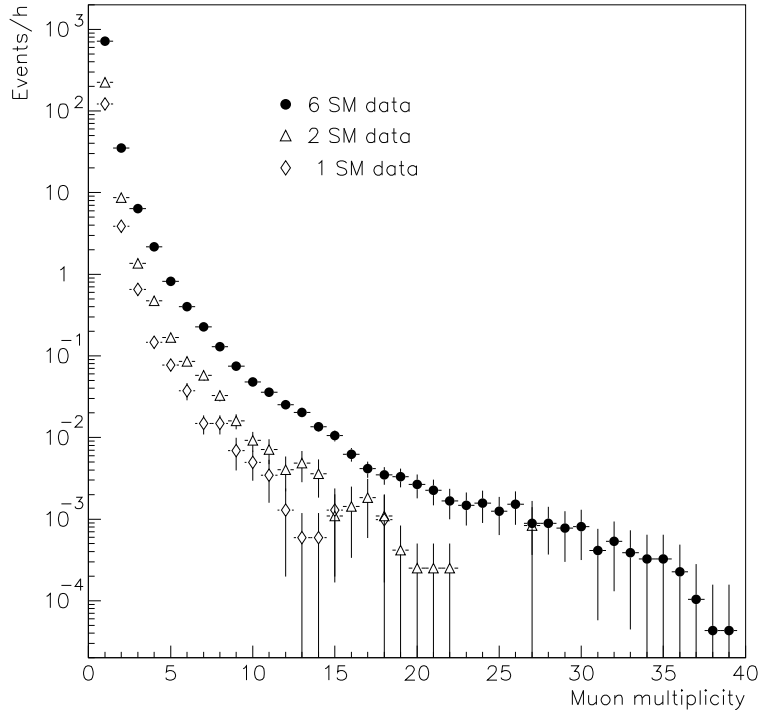

Figure 1. Muon bundle rates as functions of muon multiplicity for the one, two, and six supermodule data samples.

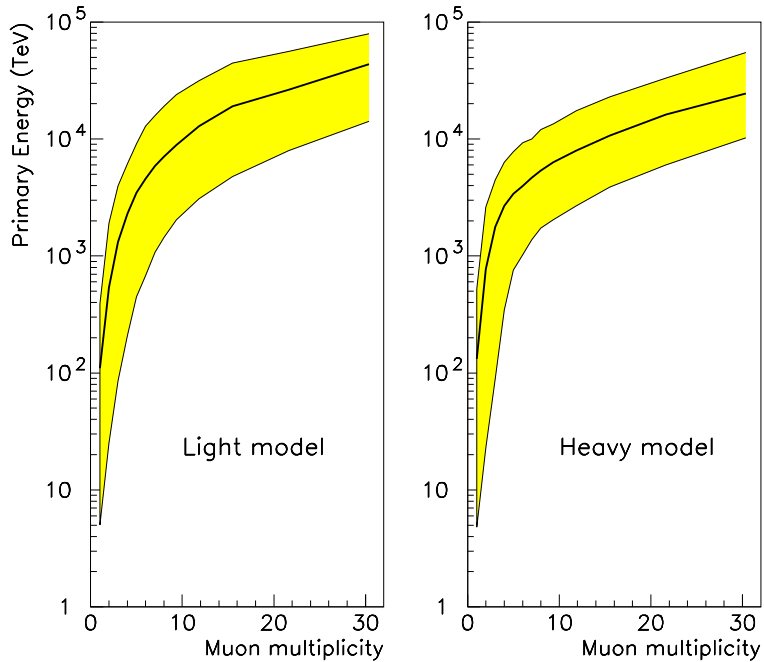

Figure 2. Calculated energy ranges of the primary cosmic rays at a detection efficiency of 90% as functions of detected multiplicities. The bold curves indicate the mean primary energies.

Figure 3 shows the dependence of the average primary mass $\langle A \rangle$ on energy for the three composition models described above, as well as for the SIGMA model which is based on fits to direct measurements from 1-100 TeV and extrapolated up to the "knee" region [11]. Therefore the SIGMA model can be viewed as a reference $\langle A \rangle$, following the energy dependence of direct measurements at low energy. As one can see from this figure, the light and CMC models are in the same range of $\langle A \rangle$ as the direct measurements. The heavy model, on the other hand, has a completely different energy dependence in nearly the entire energy range relevant to MACRO multi-muon events.

The ratios between the rates predicted by the HEMAS simulations and the experimental data are shown in Figure 4. One feature of this figure is that the measured multi-muon rates at low multiplicities ($N_\mu \leq 4$) are higher than those predicted by the Monte Carlo regardless of composition model. These events originate in primaries with energies less than a few hundred TeV, for which the three models are very similar and in agreement with direct measurements. We investigated the effects of our present uncertainties in the rock overburden and muon propagation as possible sources of this disagreement, and found from Monte Carlo simulations that while these uncertainties affect the absolute muon rates they do not affect the shape of the multiplicity distribution. An analysis based on this shape is in progress.

Figure 4 shows that our data are inconsistent with the predictions of the heavy composition at high multiplicity and favor a lighter model. Therefore our data do not favor the hypothesis used in the heavy model which requires a dramatic change of primary composition toward the pure iron element immediately above the "knee." The data provide a better fit to models with flat or slowly increasing $\langle A \rangle$ as a function of primary energy, as in the light and CMC models.

## 4. MUON DECOHERENCE

The lateral separation of underground muon pairs has been demonstrated to be sensitive to hadronic interaction models, as well as to the primary composition models, allowing the rejection of some simplified cascade treatments [12]. Here we report an analysis of a data sample of $\sim 5.8 \times 10^6$ muons and $\sim 7600$ hr live time, in which $1.9 \times 10^5$ muon pairs are reconstructed using the criteria described in [13].

The muon pair lateral separations are traditionally analyzed using a detector-independent "decoherence function" [14] defined as the rate of muon pairs per unit area, per steradian, per pair separation determined on a plane orthogonal to the pair direction:

$$G(r) = \frac{1}{\Omega T} \int \frac{d^2 N_p(r,\theta,\phi)/dr d\Omega}{A(\theta,\phi)} d\Omega \qquad (1)$$

where $d^2 N_p/dr d\Omega$ is the density of pairs at distance r and incidence angle $(\theta, \phi)$, A is the projected detector area in the $(\theta, \phi)$ direction, $\Omega$ is the total solid angle defined by the limits of integration, and $T$ is the exposure time. In this analysis, the two independent methods described in [12] are used to compute the decoherence function and they yield the same results.

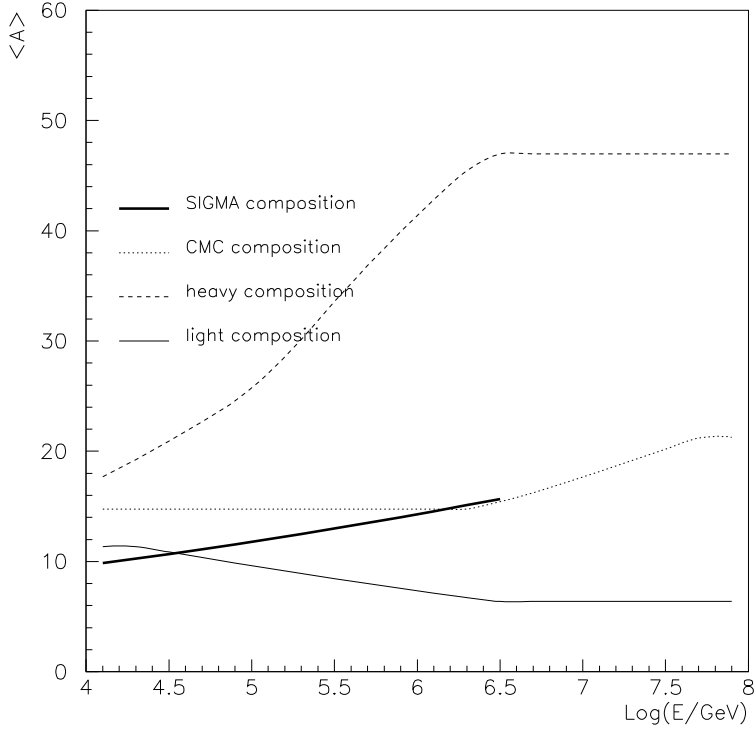

Figure 3. Average primary mass as a function of energy for various composition models.

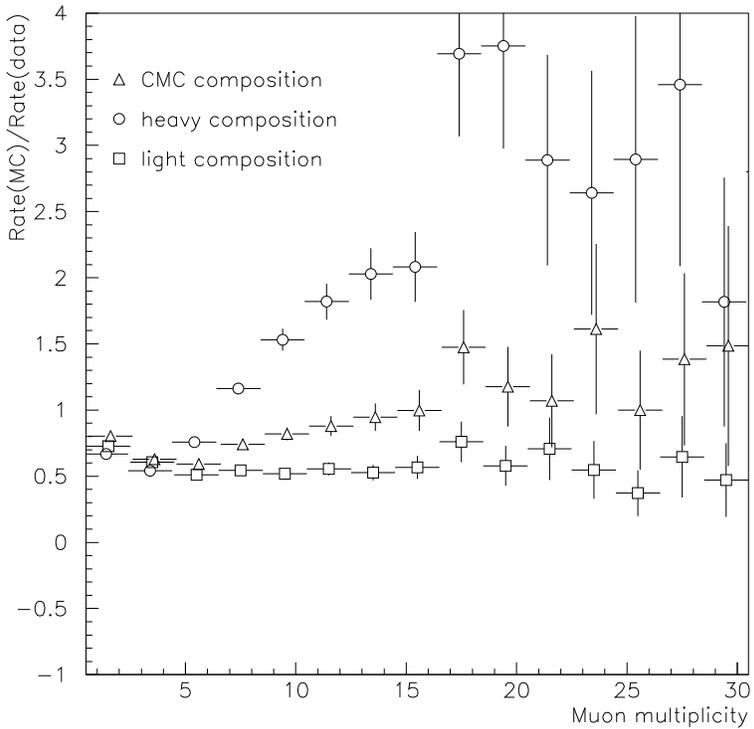

Figure 4. Ratio of predicted to observed event rates for the light, heavy, and constant mass (CMC) compositions.

To compare with composition and interaction models, a detailed Monte Carlo calculation is required to simulate the production and propagation of muons through the atmosphere and mountain overburden. The previous work [12] used the parameterized results of the HEMAS code [7] to generate cosmic ray showers, in order to save computer time. The parameterized formulae gives the number and the spatial distribution of the underground muons as functions of primary mass, energy, and direction, but it ignores the correlation between muon multiplicity and lateral distribution. Therefore we have chosen in this new analysis to perform full simulations of the atmospheric cascade and muon propagation. In preparing the full HEMAS based Monte Carlo, we investigated several possible systematic effects on the simulation, including: (1) biases due to the failure of the parameterization to account for the correlation between muon multiplicity and lateral distribution; (2) uncertainties in the primary interaction cross section; (3) different possible models of nucleus-nucleus interactions (superposition vs. fragmentation); (4) the treatment of energy loss and multiple scattering in the rock; and (5) the effect of the geomagnetic field on cascade development.

The results generally showed that the dependence of the underground muon separation on the details of the Monte Carlo was weak. A comparison of the full HEMAS based Monte Carlo to the parameterized version using the constant mass composition [10], for instance, showed that the average muon pair separation increases by only 2% in the more complete version. A 10% increase in the primary interaction cross section raises the muon production height by 3%, and consequently increase the average underground muon separation. There is a great deal of uncertainty in the nucleus-nucleus interaction model, due to the lack of accelerator experimental data for the energies of interest. The HEMAS code handles cascades generated by heavy nuclei (mass $A$, total energy $E$) in the superposition scheme, as $A$ independent nucleons of energy $E/A$. Replacing this with a more realistic model [8] causes larger fluctuations in shower development, but the decoherence distribution is not affected within the present statistics of the simulated data. HEMAS muon transport through the rock was compared to that implemented in GEANT [6], a more recent simulation developed to model high energy accelerator events; we found no noticeable difference as far as the lateral spread of muons is concerned. Finally, an increase in the average muon separation of about 5% is achieved by considering the effect of the geomagnetic field on shower development.

In Figure 5 the experimental pair separation distribution for the entire data sample is compared with Monte Carlo predictions for two extreme (light and heavy) composition models. In the previous analysis [12] of data from only two supermodules, we found a good agreement with the Monte Carlo predictions up to the maximum attainable separation ($\sim 20$ m). With all six lower supermodules the measured decoherence function at large separations is higher than that from the simulations. The average separation is 10.9±0.2 m for the real data, 10.5 m for the heavy model, and 9.4 m for the light model. The measured separation is larger than the one from the simulations. We have also investigated the dependence of muon pair separation on rock depth and zenith angle. Tables 1 and 2 summarize the main results. It is worth noting that the experimental and simulated values follow the same behavior and are close to one another. But the experimental values are larger than the simulated ones, following the same trend in the overall distribution.

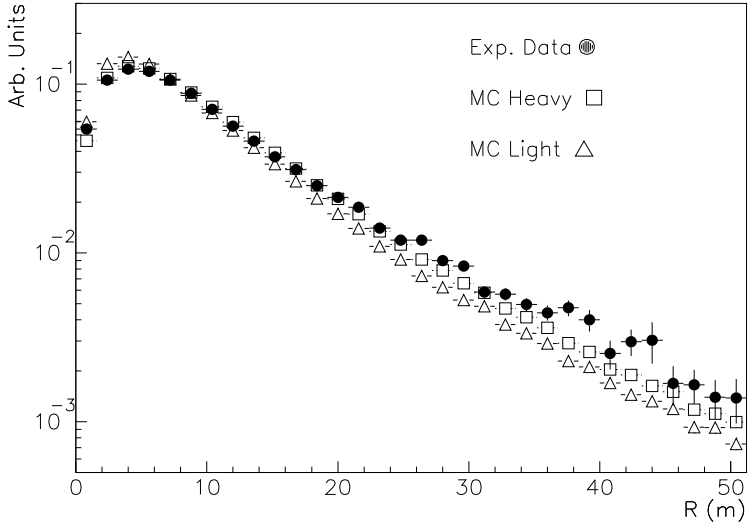

Figure 5. The decoherence function as a function of the muon pair separation.

Since the heavy model is a disfavored extreme model (see Section 3), Figure 5 may indicate that the shower development is not yet treated properly in the Monte Carlo and that the hadronic interaction model presently used for this analysis may need further improvement. In particular, we are investigating factors which affect the transverse momentum distribution, including possible nuclear effects.

In order to better understand the role of the hadronic interaction model, we have performed a comparative simulation with the SIBYLL code [9]. No substantial changes, however, were observed as far as muon separation is concerned. Such a model in fact predicts a slightly lower average muon separation (on the order of 10% less). In the future, other models will be considered, such as the DPMJET code [15], which has a more complete treatment of nuclear effects than does SIBYLL.

## 5. MUON DECORRELATION

Detailed measurements of quantities relative to underground multiple muon events can provide information on primary interactions and muon propagation through the rock overburden. In particular, the differential distribution of spatial and angular separations ($x$ and $\phi$) in multiple muon events, $dN/dxd\phi$, is sensitive to the physics of muon production and propagation. It provides information on the total primary cross section as well as to the transverse momentum distribution of the parent hadrons of the underground muons. It also provides a measure of the effects of muon interactions in the rock overburden, which introduce displacements of the muons from their original direction and position. The aim of this analysis is to attempt to disentangle these effects.

9Table 1
Comparison of experimental lateral separations with MC results for different zenith angle intervals (depth = 3750 → 4150 m.w.e.). $\langle D \rangle$ is the average separation and $R_0$ is the separation at which the decoherence function reaches its maximum. Both quantities are in unit of meters.

| $\cos\theta$ | exp. $R_0$ | exp. $\langle D \rangle$ | $\sigma_{\text{stat}}$ $R_0$ | $\sigma_{\text{stat}}$ $\langle D \rangle$ | $\sigma_{\text{syst}}$ $\langle D \rangle$ | MC Light $R_0$ | MC Light $\langle D \rangle$ | MC Heavy $R_0$ | MC Heavy $\langle D \rangle$ |
|---|---|---|---|---|---|---|---|---|---|
| 0.5 → 0.6 | 6.50 | 14.8 | 0.10 | 0.2 | 0.4 | 5.70 | 13.3 | 6.70 | 14.8 |
| 0.6 → 0.7 | 5.29 | 13.4 | 0.05 | 0.1 | 0.5 | 4.53 | 11.1 | 5.35 | 12.3 |
| 0.7 → 0.8 | 4.52 | 12.5 | 0.04 | 0.1 | 0.7 | 3.89 | 9.2 | 4.53 | 10.3 |
| 0.8 → 0.9 | 3.91 | 9.7 | 0.05 | 0.1 | 0.3 | 3.37 | 8.1 | 3.82 | 8.9 |
| 0.9 → 1.0 | 3.43 | 7.7 | 0.04 | 0.1 | 0.2 | 2.99 | 7.0 | 3.44 | 7.9 |

Table 2
Comparison of experimental lateral separation with MC results for different rock depth intervals ($\cos\theta = 0.8 \to 0.9$). $\langle D \rangle$ is the average separation and $R_0$ is the separation at which the decoherence function reaches its maximum. Both quantities are in unit of meters.

| Rock Depth (m.w.e) | Exp. $R_0$ | Exp. $\langle D \rangle$ | $\sigma_{\text{stat}}$ $R_0$ | $\sigma_{\text{stat}}$ $\langle D \rangle$ | $\sigma_{\text{syst}}$ $\langle D \rangle$ | MC Light $R_0$ | MC Light $\langle D \rangle$ | MC Heavy $R_0$ | MC Heavy $\langle D \rangle$ |
|---|---|---|---|---|---|---|---|---|---|
| 3350 → 3750 | 4.29 | 10.3 | 0.04 | 0.1 | 0.2 | 3.88 | 9.1 | 4.44 | 10.1 |
| 3750 → 4150 | 3.91 | 9.7 | 0.05 | 0.1 | 0.3 | 3.37 | 8.1 | 3.82 | 8.9 |
| 4150 → 4550 | 3.03 | 6.8 | 0.07 | 0.1 | 0.2 | 2.76 | 6.6 | 3.14 | 7.4 |
| 4550 → 4950 | 2.63 | 6.6 | 0.07 | 0.2 | 0.3 | 2.44 | 6.1 | 2.86 | 6.9 |

For a real experiment of finite size and finite live time, it is generally impossible to reconstruct the full distribution, $dN/dxd\phi$. Grillo and Parlati [16] have suggested the use of its first moment,

$$\langle \phi(x) \rangle = \frac{\int \phi \frac{dN}{dxd\phi} d\phi}{\int \frac{dN}{dxd\phi} d\phi}, \qquad (2)$$

which they have named as the "decorrelation function." An analytic expression for $\langle \phi(x) \rangle$ has been derived [16] using some approximations.

In this analysis, we have used only double muon events and employed track reconstruction only in one projective view, namely the wire view. The results are still preliminary. Since the position of the shower axis is not known, the measure is relative; that is, the quantities $\phi$ and $x$ are defined between pairs of muons in the same bundle. The distance is always taken to be positive, while the angle is positive if the tracks are diverging, and negative otherwise. Since the average angle as a function of distance is normalized to the number of events at that distance, the influence of apparatus effects (efficiency, working conditions, containment) should not be important. We have made no run selection on the data sample, but we have made an event selection rejecting too short tracks, to avoid contamination of the sample by locally produced pions and small showers. The experimentally measured decorrelation function is presented in Figure 6.

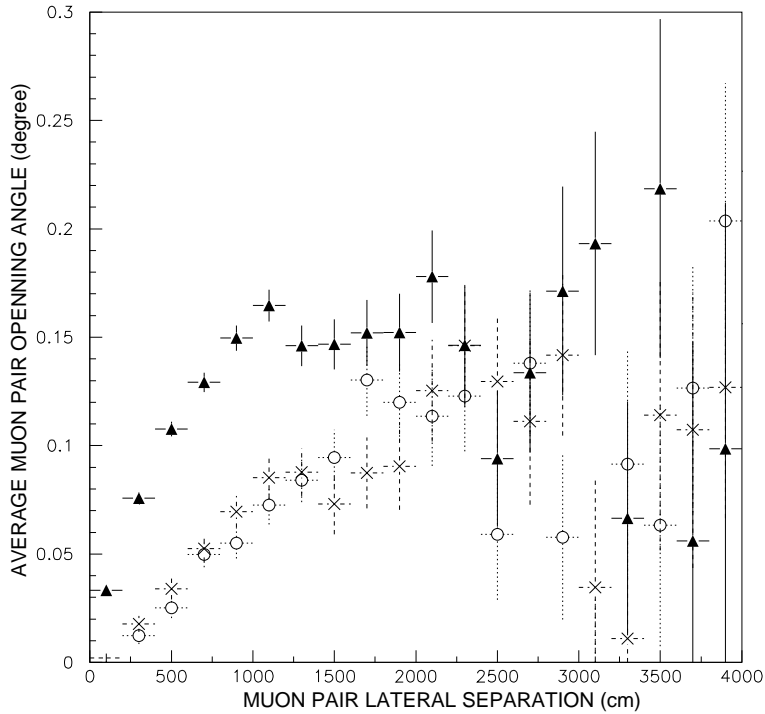

Figure 6. The decorrelation function as measured in MACRO (triangles) compared with the HEMAS predictions for the extreme light (crosses) and heavy (circles) composition models.

Figure 6 also shows the decorrelation function computed from the HEMAS code. It is evident that there is a strong disagreement between the data and Monte Carlo at relatively small distances. We have investigated several possible causes for this disagreement. Different composition models have essentially no effect, nor do more refined muon transport codes which include Molière tails beyond the Gaussian approximation for multiple scattering. It is possible to modify the average interaction cross section (and hence primary interaction height) to make the Monte Carlo results agree with the data, but the required modification is rather extreme and is likely inconsistent with reasonable extrapolations of accelerator data. We are presently investigating more subtle effects, both derived from the finite space resolution of the apparatus and from high energy interactions of muons in the rock overburden. Preliminary results are encouraging, but a more refined analysis is not yet complete.

## 6. CONCLUSIONS

The large amount of underground muon bundle data collected by the MACRO experiment offer significant capability in the studies of primary cosmic ray composition and of interactions at very high energies. Our data do not favor the hypothesis of a dramatic change of primary composition towards pure Fe element immediately above the "knee." The discrepancy between the experimental muon pair lateral separations and the Monte Carlo results is under investigation. A new analysis based on the muon decorrelation function is being pursued.

We are very grateful to the U.S. Department of Energy, the Laboratori Nazionali del Gran Sasso, and the Istituto Nazionale di Fisica Nucleare for their continued support of the MACRO experiment.

# † The MACRO Collaboration

S. Ahlen[3], M. Ambrosio[12], R. Antolini[7], G. Auriemma[14,a], R. Baker[11], A. Baldini[13], G. C. Barbarino[12], B. C. Barish[4], G. Battistoni[6,19★], R. Bellotti[1], C. Bemporad[13], P. Bernardini[10], H. Bilokon[6], V. Bisi[16], C. Bloise[6], C. Bower[8], S. Bussino[14], F. Cafagna[1], M. Calicchio[1], D. Campana[6], M. Carboni[6], S. Cecchini[2,b], F. Cei[13], V. Chiarella[6], R. Cormack[3], A. Corona[14], S. Coutu[11], G. DeCataldo[1], H. Dekhissi[2,c], C. DeMarzo[1], I. De Mitri[9], M. De Vincenzi[14,d], A. Di Credico[7], E. Diehl[11], O. Erriquez[1], C. Favuzzi[1], D. Ficenec[3,e], C. Forti[6], P. Fusco[1], G. Giacomelli[2], G. Giannini[13,f], N. Giglietto[1], M. Goretti[14], M. Grassi[13], P. Green[15,18★], A. Grillo[6], F. Guarino[12], P. Guarnaccia[1], C. Gustavino[7], A. Habig[8], K. Hanson[11], R. Heinz[8], J. T. Hong[3], E. Iarocci[6,g], E. Katsavounidis[4], E. Kearns[3], S. Klein[3,h], S. Kyriazopoulou[4], E. Lamanna[14], C. Lane[5], D. S. Levin[11], P. Lipari[14], G. Liu[4], R. Liu[4], M. J. Longo[11], Y. Lu[15], G. Ludlam[3], G. Mancarella[10], G. Mandrioli[2], A. Margiotta-Neri[2], A. Marin[3], A. Marini[6], D. Martello[10,17★], A. Marzari Chiesa[16], P. Matteuzzi[2], M. N. Mazziotta[1], D. G. Michael[4], S. Mikheyev[7,i], L. Miller[8], M. Mittelbrun[5], P. Monacelli[9], T. Montaruli[1], M. Monteno[16], S. Mufson[8], J. Musser[8], D. Nicoló[13], R. Nolty[4] S. Nutter[11], C. Okada[3], G. Osteria[12], O. Palamara[10], S. Parlati[4,7★], V. Patera[6,g], L. Patrizii[2], B. Pavesi[2], R. Pazzi[13], C. W. Peck[4], J. Petrakis[8,17★], S. Petrera[10], N. D. Pignatano[4], P. Pistilli[10,14★], A. Rainó[1], J. Reynoldson[7], F. Ronga[6], G. Sanzani[2], A. Sanzgiri[15], F. Sartogo[14], C. Satriano[14,a], L. Satta[6,g], E. Scapparone[2], K. Scholberg[4], A. Sciubba[6,g], P. Serra Lugaresi[2], M. Severi[14], M. Sitta[16], P. Spinelli[1], M. Spinetti[6], M. Spurio[2], J. Steele[5], R. Steinberg[5], J. L. Stone[3], L.R. Sulak[3], A. Surdo[10], G. Tarlé[11], V. Togo[2], V. Valente[6], C. W. Walter[4], R. Webb[15], and W. Worstell[3]

1. Dipartimento di Fisica dell'Università di Bari and INFN, Bari, 70126, Italy
2. Dipartimento di Fisica dell'Università di Bologna and INFN, Bologna, 40126, Italy
3. Physics Department, Boston University, Boston, MA 02215, USA
4. California Institute of Technology, Pasadena, CA 91125, USA
5. Department of Physics, Drexel University, Philadelphia, PA 19104, USA
6. Laboratori Nazionali di Frascati dell'INFN, Frascati (Roma), 00044, Italy
7. Laboratori Nazionali del Gran Sasso dell'INFN, Assergi (L'Aquila), 67010, Italy
8. Depts. of Physics and of Astronomy, Indiana University, Bloomington, IN 47405, USA
9. Dipartimento di Fisica dell'Università dell'Aquila and INFN, L'Aquila, 67100, Italy
10. Dipartimento di Fisica dell'Università di Lecce and INFN, Lecce, 73100, Italy
11. Department of Physics, University of Michigan, Ann Arbor, MI 48109, USA
12. Dipartimento di Fisica dell'Università di Napoli and INFN, Napoli, 80125, Italy
13. Dipartimento di Fisica dell'Università di Pisa and INFN, Pisa, 56010, Italy
14. Dipartimento di Fisica dell'Università di Roma and INFN, Roma, 00185, Italy
15. Physics Department, Texas A&M University, College Station, TX 77843, USA
16. Dipartimento di Fisica dell'Università di Torino and INFN, Torino, 10125, Italy
17. Bartol Research Institute, University of Delaware, Newark, DE 19716, USA
18. Sandia National Laboratory, Albuquerque, NM 87185, USA
19. INFN Sezione di Milano, 20133, Italy

★ Current address
a Also Università della Basilicata, Potenza, 85100, Italy
b Also Istituto TESRE/CNR, Bologna, Italy
c Also at Faculty of Science, University Mohamed I, Oujda, Morocco
d Also at Università di Camerino, Camerino, Italy
e Now at Physics Department, Washington University, St Louis, MO 63130, USA
f Also at Università di Trieste and INFN, Trieste, 34100, Italy
g Also at Dipartimento di Energetica, Università di Roma, Roma, 00185, Italy
h Now at Department of Physics, University of California, Santa Cruz, CA 95064, USA
i Also at Institute for Nuclear Research, Russian Academy of Science, Moscow, Russia